\documentclass[entropy,article,
submit,
moreauthors,pdftex,10pt,a4paper]{mdpi} 

\firstpage{1} 
\articlenumber{x}
\doinum{10.3390/------}
\pubvolume{xx}
\pubyear{2018}
\copyrightyear{2018}
\history{Received: date; Accepted: date; Published: date}
\preto{\abstractkeywords}{\nolinenumbers}

\usepackage{amsmath,amssymb}

\newcommand{\Fo}{{\rm F}_{\rm o}}
\def\F1{{\rm F}_{1}}
\newcommand{\mem}{I_\mathrm{mem}}
\newcommand{\pred}{I_\mathrm{pred}}
\newcommand{\Nos}{I_\mathrm{nos}}
\newcommand{\NosSS}{\Nos^\mathrm{ss}}
\newcommand{\phiSS}{\phi^{\mathrm{ss}}}
\newcommand{\kenv}{\kappa_\mathrm{env}}
\newcommand{\kenvAv}{\bar{\kappa}_\mathrm{env}}
\newcommand{\ksys}{k_\mathrm{sys}}
\newcommand{\WdissSS}{\beta \left\langle W_{\mathrm{diss}}^{ss} \right\rangle}
\newcommand{\kstar}{\ksys/\kenv}

\newcommand*\xbar[1]{%
   \hbox{%
     \vbox{%
       \hrule height 0.5pt 
       \kern0.4ex
       \hbox{%
         \kern-0.1em
         \ensuremath{#1}%
         \kern-0.1em
       }%
     }%
   }%
} 

\Title{Energy Dissipation and Information Flow in Coupled Markovian Systems}

\Author{Matthew E.~Quenneville $^{1,2,}$* and David A.~Sivak $^{1,}$*}
\AuthorNames{Matthew E.~Quenneville and David A.~Sivak}

\address{%
$^{1}$ \quad Department of Physics, Simon Fraser University, Burnaby, British Columbia, Canada\\
$^{2}$ \quad Department of Physics, University of California, Berkeley, California, USA\\}

\corres{Correspondence: mquenneville@berkeley.edu, dsivak@sfu.ca}

\abstract{
A stochastic system under the influence of a stochastic environment is correlated with both present and future states of the environment. 
Such a system can be seen as implicitly implementing a predictive model of future environmental states. The non-predictive model complexity 
has been shown to lower-bound
the
thermodynamic dissipation. 
Here we explore 
these statistical and physical quantities at steady state in simple models. 
We show that 
under quasi-static driving
this model complexity 
saturates
the dissipation.
Beyond the quasi-static limit, 
we demonstrate 
a lower bound on the ratio of this model complexity to total dissipation, that is realized 
in the limit of weak driving. 
}

\keyword{work; dissipation; quasi-static; information; prediction; learning; nostalgia. 
}

\begin{document}




\section{Introduction}

Information theory has long been recognized as fundamentally linked to statistical mechanics~\cite{Jaynes:1957ua}. Perhaps most prominently, Landauer showed that information processing can require unavoidable dissipative costs~\cite{landauer1961irreversibility}; for example, bit erasure requires that some free energy be dissipated~\cite{Berut2012,Jun2014}. 

A stochastic system processes information through interaction with its environment: 
through environment-dependent dynamics the system responds to environmental changes and thereby gains information about the environment~\cite{Cheong:2011jp,Mehta:2012ji}.
For an 
environment exhibiting temporal correlations, the system carries information about the past, present, and future environmental states.
In this way, the system implicitly implements a predictive model of future environmental states~\cite{Still2012}.

One can
quantify this model's inefficiency
by
the unnecessary model complexity: 
information the model retains about the past 
that does not aid in predicting the future. Recent work established the equivalence between this predictive inefficiency and thermodynamic inefficiency~\cite{Still2012}, providing another fundamental connection between information theory and statistical mechanics. This connection hints at 
a design principle for 
molecular
machines operating out of equilibrium~\cite{Hess:2011:AnnuRevBiomedEng,brown2017toward}. 

These results are potentially applicable to many systems. For example, biological molecular machines generally operate far from equilibrium within highly stochastic environments. ATP synthase, a molecular machine which synthesizes adenosine triphosphate (ATP), is composed of two sub-units. The first ($\Fo$) drives the second sub-unit ($\F1$), which in turn produces ATP. The crankshaft rotation of $\Fo$ that mechanically drives $\F1$ is stochastic. In this way, $\F1$ contains an implicit prediction of future rotations of $\Fo$. In order for ATP synthesis to proceed at minimum energetic cost, the implicit model should contain little extraneous model complexity~\cite{okuno2011rotation}.
Examples like this occur throughout 
biology with organisms~\cite{Tagkopoulos:2008ct}, neurons~\cite{Laughlin:1981wn}, reaction networks~\cite{McGregor:2012ij}, and even potentially single molecules learning statistical patterns in their respective environments.

To further illuminate this abstract connection between model complexity and thermodynamic dissipation, 
here
we analytically and numerically explore these statistical and physical quantities in illustrative models. 
We demonstrate the information learned by the system about its environment per unit energy dissipated (equivalently the ratio of dissipation during system and environmental dynamics) in the limits of quasi-static driving (Table~\ref{tab:bounds}) and weak driving~\eqref{eq:limit_smallprobs}, which forms the lower bound for generic driving. 
The dependence of these quantities on the 
system and environmental 
parameters 
motivates
a potential guiding principle 
for functional performance. 

\section{Theoretical Background}
\begin{figure}[H]
\centering
\includegraphics[width=0.7\textwidth]{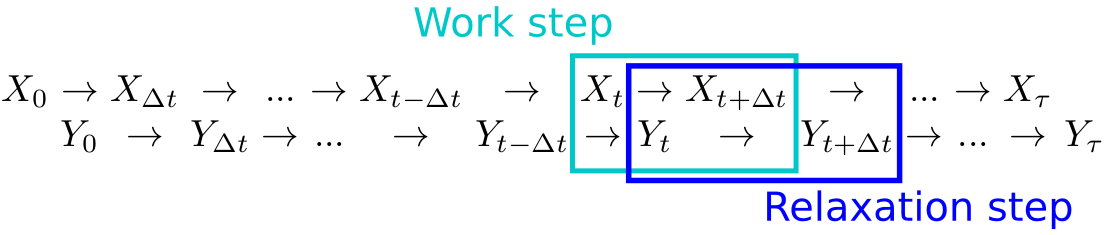}
\caption{{\bf Discrete-time system and environmental dynamics.} 
The system $Y_t$ and environment $X_t$ alternate steps, with system evolution during relaxation steps, and environment evolution during work steps.}
\label{fig:timestep}
\end{figure}

Consider a stochastic process $\{ X_{t} | t \in \{0,\Delta t, ..., \tau-\Delta t, \tau\}\}$ representing the dynamical evolution of some environmental variable. At a given time, the environment can occupy any of the states $\mathcal{X}$. The time evolution of the environment, $X_{t}$, is governed by the transition probabilities $p(x_t|\{ x_{t'} \}_{t'=0}^{t-\Delta t}) \equiv p(X_{t}=x_t| \{ X_{t'} = x_{t'} \}_{t'=0}^{t-\Delta t})$ for $x_t,x_{t'} \in \mathcal{X}$. 
Let another stochastic process $\{ Y_{t} | t \in \{0,\Delta t, ..., \tau-\Delta t, \tau\}\}$ represent the system of interest, which can occupy states $\mathcal{Y}$. Take the dynamics of $Y_{t}$ to depend on the environmental state via the time-independent conditional transition probabilities $p(y|y',x) \equiv p(Y_{t+\Delta t}=y|Y_{t}=y',X_{t+\Delta t}=x)$, where $y, y' \in \mathcal{Y}$.
We model the evolution of these two stochastic processes using an alternating time-step pattern illustrated in Fig.~\ref{fig:timestep}. One complete time-step is composed of two sub-steps: one work step of environmental dynamics, when the environment does work on the system, followed by one relaxation step of system dynamics, when the system exchanges heat with a thermal bath maintained at temperature $T$ and inverse temperature $\beta \equiv (k_{\rm B}T)^{-1}$. 

System dynamics $Y_{t}$ obey the principle of microscopic reversibility~\cite{Chandler1987a}. Ref.~\cite{Still2012} used such a framework to study the relationship between thermodynamic and information-theoretic quantities. One prominent information-theoretic quantity is the \emph{nostalgia} $\Nos(t) \equiv \mem(t)-\pred(t)$, where the mutual information $\mem(t) \equiv I[X_{t},Y_{t}]$ \cite{cover2012elements} between the current system state and past environmental state represents the memory stored by the system about the environment, and the mutual information $\pred(t) \equiv I[X_{t+\Delta t},Y_{t}]$ between current system state and future environmental state represents the ability of the system to predict future environmental states. Ref.~\cite{Still2012} showed that
\begin{equation}
  \label{eq:ToP}
    \beta \left\langle W_{\mathrm{diss}}(t) \right\rangle = \mem(t) - \pred(t) - \beta \left\langle \Delta F_{\mathrm{neq}}^{\mathrm{relax}}(t) \right\rangle \ ,
\end{equation}
where $\left\langle W_{\mathrm{diss}}(t) \right\rangle$ is the total dissipation over the step from $t$ to $t+\Delta t$, and $\left\langle \Delta F_{\mathrm{neq}}^{\mathrm{relax}}(t) \right\rangle$ is the change in (nonequilibrium) free energy over the relaxation step from $t$ to $t+\Delta t$.
Since $\beta \left\langle \Delta F_{\mathrm{neq}}^{\mathrm{relax}}(t) \right\rangle \le 0$~\cite{Schnakenberg:1976p55305},
\begin{equation}
  \label{eq:ToPinequality}
  \beta \left\langle W_{\mathrm{diss}}(t) \right\rangle \ge \mem(t) - \pred(t) \ .
\end{equation}

\section{Results}

We explore the tightness of the bound~\eqref{eq:ToPinequality} through the ratio of nostalgia to dissipation,
\begin{equation}
\label{eq:phi}
\phi(t) \equiv \frac{\mem(t) - \pred(t)}{\beta \left\langle W_{\mathrm{diss}}(t) \right\rangle} \ .
\end{equation}
This nostalgia-dissipation ratio is bounded by $0 \le \phi(t) \le 1$ and (after substituting Eq.~(14) from \cite{Still2012}) can be interpreted as the fraction of dissipation which occurs over work steps,
\begin{equation}
\label{eq:phiwork}
\phi(t) = \frac{\left\langle W_{\mathrm{diss}}[x_t \rightarrow x_{t+\Delta t}] \right\rangle}{\left\langle W_{\mathrm{diss}}(t) \right\rangle} \ .
\end{equation}

When the environment and system reach steady state, $\phi$ can be rewritten as:
\begin{equation}
\label{eq:phi_ell}
\phiSS = \frac{\ell(t)}{-\beta \left\langle Q \right\rangle},
\end{equation}
where $\ell(t) \equiv I[X_{t+\Delta t},Y_{t+\Delta t}]-I[X_{t+\Delta t},Y_{t}]$ is a learning rate which quantifies the information gained by the system about the current environmental state~\cite{Brittain2017}. 
The denominator follows from the facts that at steady state $-\left\langle Q \right\rangle= \left\langle W \right\rangle$ (due to energy conservation) 
and
$\left\langle W \right\rangle=\left\langle W_\mathrm{diss} \right\rangle$~\cite{Still2012}.
\cite{Barato2014} and \cite{hartich2016sensory} identify the ratio in Eq.~\eqref{eq:phi_ell} as an informational efficiency quantifying the rate at which the system learns about the environment, relative to the total thermodynamic entropy production. 
By considering \eqref{eq:phiwork}, these results can be recast in terms of dissipative energy flows.

In order to explore the physical implications of \eqref{eq:ToP} and \eqref{eq:ToPinequality}, we investigate the behavior of the relevant information-theoretic and thermodynamic quantities in concrete models that provide physical intuition. We initially restrict our attention to a simple environment model, consisting of two states with a constant transition rate $\kenv$.

\subsection{Alternating Energy Levels}
\label{sec:antisym}
\begin{figure}[H]
\centering
\includegraphics[width=\textwidth]{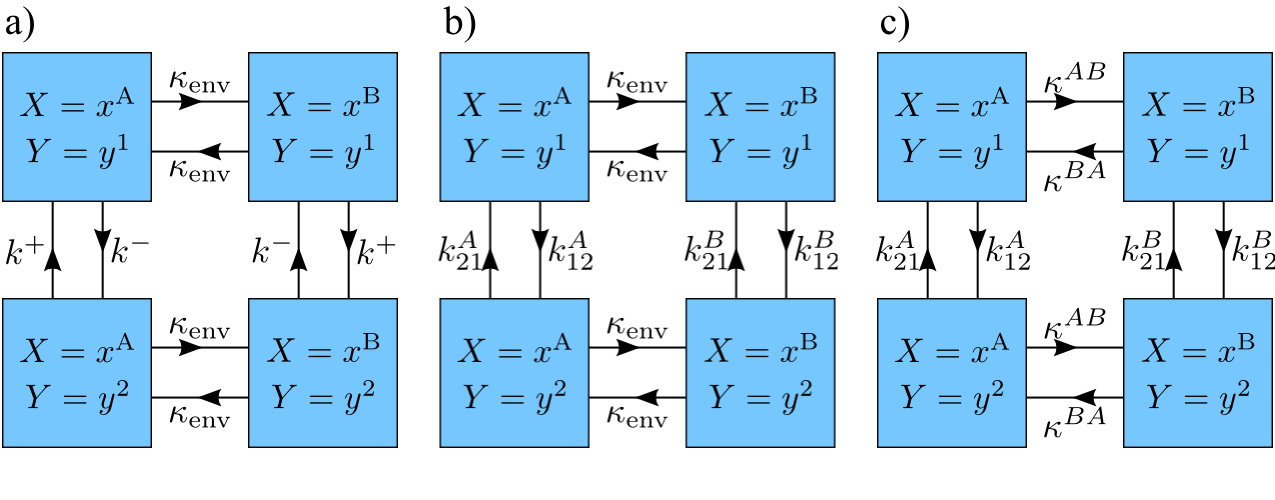}
\caption{{\bf Model kinetics.} 
States and transition rates for models with two system states and two environment states. (\textbf{a}) System equilibration rate and energy gap magnitude and environment transition rate are independent of environment state, but the direction of the energy gap switches with environment state. (\textbf{b}) System equilibration rate and energy gap vary with environment state. Environment transition rate is fixed. (\textbf{c}) System equilibration rate and energy gap and environment transition rate vary with environment state. 
}
\label{fig:statemaps}
\end{figure}

One of the simplest possible system models with non-trivial behavior is a two-state system with dynamics described by two kinetic rates, $k_{+}$ and $k_{-}$ (Fig.~\ref{fig:statemaps}a). 
This model possesses a symmetry such that it is unchanged when both the system-state labels and environment-state labels are interchanged. Due to this symmetry, we take $k_{+} \ge k_{-}$ without loss of generality. 

\begin{figure}[H]
\centering
\includegraphics[width=0.56\textwidth]{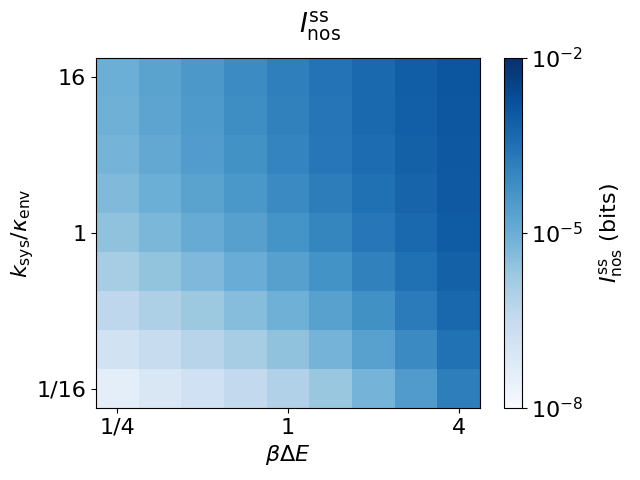}
\caption{{\bf Nostalgia increases with energy gap and system equilibration rate.} Nostalgia $\NosSS$ as a function of the energy gap $\beta \Delta E$ and transition ratio $\kstar$. ($\kenv \Delta t = 10^{-12}$.)}
\label{fig:nos_hm}
\end{figure}

Given the constraint of detailed balance~\cite{Chandler1987a}, such a model describes a two-state system with an energy gap (normalized by temperature) $\beta \Delta E = \ln \frac{k_{+}}{k_{-}}$ that flips according to the environment state. 
System states $y^1$ and $y^2$ are separated by $\Delta E^{A}_{12}=-\Delta E$ when the environment is in state $x^A$ and $\Delta E^{B}_{12}=\Delta E$ for environmental 
state $x^B$. The characteristic rate at which the system equilibrates, and thus becomes correlated with the current environment (and decorrelated with past environmental states), is the harmonic mean of the two transition rates, 
\begin{equation}
\ksys \equiv \frac{2}{\frac{1}{k_{+}} + \frac{1}{k_{-}}} \ . 
\end{equation}
The transition ratio $\kstar$ expresses this rate relative to the environmental transition rate.
Figure~\ref{fig:nos_hm} shows the steady-state nostalgia $\NosSS$, which increases with both $\kstar$ and $\beta \Delta E$, and tends to 0 as either $\kstar$ or $\beta \Delta E$ approach 0.

\begin{figure}[H]
\centering
\includegraphics[width=0.56\textwidth]{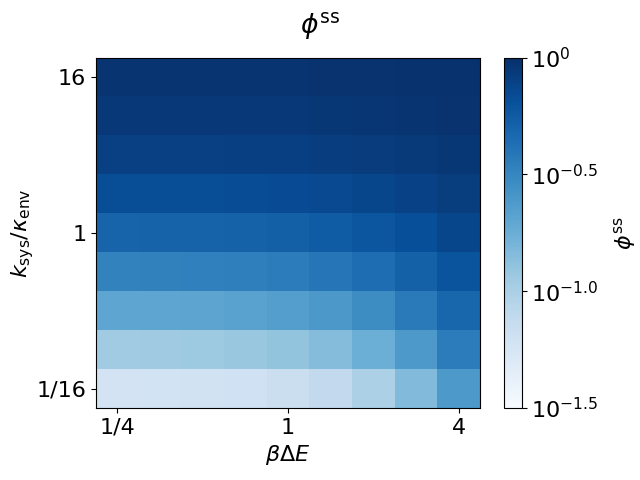}
\caption{{\bf 
Dissipation ratio increases with energy gap and system equilibration rate.} Steady-state dissipation ratio $\phiSS\equiv\NosSS/\WdissSS$ as a function of the energy gap $\beta \Delta E$ and transition ratio $\kstar$. ($\kenv \Delta t = 10^{-12}$.)}
\label{fig:phi_hm}
\end{figure}

The dissipation ratio $\phi(t)$ approaches a steady-state value $\phiSS$ for each choice of parameters. Figure~\ref{fig:phi_hm} shows that $\phiSS$ follows the same general trends as $\NosSS$, increasing with both energy gap magnitude $\beta \Delta E$ and transition ratio $\kstar$.

In the limit of large temperature, when the energy gap is small compared to the ambient thermal energy ($\beta \Delta E \ll 1$), $\phiSS$ reduces to a positive function of the equilibration rates of the system ($\ksys$) and environment ($\kenv$):
\begin{equation}
	\label{eq:limit}
	\phiSS = \frac{1-\kenv\Delta t}{1-2\kenv\Delta t+\kenv/\ksys} , \quad \beta\Delta E \ll 1 \ . 
\end{equation}
This is found by explicitly calculating the steady-state probability distribution.
In moving from discrete-time steps to a continuous-time parameter, 
time step size becomes small compared to system and environment transition times,
reducing \eqref{eq:limit} to
\begin{equation}
	\label{eq:limit_smallprobs}
	\phiSS = \frac{1}{1+\kenv/\ksys} \ ,\quad \kenv\Delta t, \ksys\Delta t, \beta\Delta E \ll 1 \ .
\end{equation}

Thus, in the weak driving (high-temperature) limit ($\beta \Delta E \ll 1$), if the system evolves quickly compared to the environment, most of the dissipation occurs during work steps, the learning rate approaches the total thermodynamic entropy production, and the bound~\eqref{eq:ToPinequality} approaches saturation. 
Conversely (still restricting to high temperature), when the system evolves slowly compared to the environment, most of the dissipation occurs during relaxation steps, the learning rate is small compared to the total thermodynamic entropy production, and the nostalgia is small compared to the bound in \eqref{eq:ToPinequality}.

Further, Fig.~\ref{fig:phi_hm} shows that $\phiSS$ increases with $\beta \Delta E$. 
Thus, this weak-driving limit gives a non-zero lower bound on $\phiSS$,
\begin{equation}
	\label{eq:bound}
	\frac{1-\kenv\Delta t}{1-2\kenv\Delta t+\kenv/\ksys} \le \phiSS \le 1 \ ,
\end{equation}
or in the limit of small time steps,
\begin{equation}
	\label{eq:bound_smallprobs}
	\frac{1}{1+\kenv/\ksys} \le \phiSS \le 1 \ , \quad 
    \kenv\Delta t, \ksys\Delta t\ll 1 \ .
\end{equation}
If the system evolves quickly compared to its environment, nostalgia is the dominant form of dissipation, regardless of $\beta \Delta E$. 
The limit of quasi-static driving is defined by $\kstar \gg 1$. In this limit, $\phiSS=1$, and therefore the nostalgia (the implicit predictive model inefficiency) is equal to the total dissipation (the thermodynamic inefficiency). The bounds in Eqs.~\eqref{eq:bound} and \eqref{eq:bound_smallprobs} therefore hold beyond the quasi-static limit. The bound in Eq.~\eqref{eq:ToPinequality} can be looser for systems farther from the limit of quasi-static driving. 
These limits on $\phiSS$ are laid out in Table~\ref{tab:bounds}.


\begin{table}[H]
    \caption{    \label{tab:bounds}
{\bf Limiting behavior of dissipation ratio.} Steady-state dissipation ratio $\phiSS$ in the various limits of driving strength and speed. These limits are given by the bound in Eq.~\eqref{eq:bound_smallprobs}, valid in the limit of continuous time.}
\centering
    \begin{tabular}{cc|c|c}
    ~             & \textbf{Driving Strength}     & Weak                        & Strong                                \\
    \textbf{Driving Speed} & ~                    & ($\beta\Delta E \ll 1$)     & ($\beta\Delta E \gtrsim 1$)        \\ \hline
    Quasi-static  & ($\kenv \ll \ksys$)  & $\phiSS=1$                    & $\phiSS=1$                              \\
    Intermediate & ($\kenv \sim \ksys$) & $\phiSS=(1+\kenv/\ksys)^{-1}$ & $(1+\kenv/\ksys)^{-1} \le \phiSS \le 1$ \\
    Fast          & ($\kenv \gg \ksys$)  & $\phiSS=\ksys/\kenv$           & $\ksys/\kenv \le \phiSS \le 1$          \\ \hline
    \end{tabular}
\end{table}

The transition ratio $\ksys/\kenv$ is also equal to the ratio of characteristic timescales $\tau_{\rm env}/\tau_{\rm sys}$. Thus the bound for steady-state dissipation ratio~\eqref{eq:bound_smallprobs} can be recast as 
\begin{equation}
\frac{1}{1+N} \le \phiSS \le 1 \ , \quad 
    \kenv\Delta t, \ksys\Delta t\ll 1 \ ,
\end{equation}
for $N$ independent `measurements' the system makes during each environment state~\cite{Govern:2014ez}.  From this perspective, the bound is proportional (up to a multiplicative constant) to the Berg-Purcell lower bound on environmental measurement precision of a single receptor~\cite{Berg:1977bp}.

\subsection{Arbitrary System Rates}
The results of the previous section were derived for a simple two-state system, in which the energy difference between system states flips with environment transitions, and the system's equilibration rate is independent of the environment state. We generalize this model to a two-state system with arbitrary rates and hence---by detailed balance---arbitrary energies (Fig.~\ref{fig:statemaps}b). Given the four transition rates $k_{12}^{A}$, $k_{21}^{A}$, $k_{12}^{B}$, and $k_{21}^{B}$, when the environment is in state $X=x^{A}$ the system has energy gap (normalized by temperature) $\beta \Delta E^A_{12} = \ln \frac{k_{21}^{A}}{k_{12}^{A}}$ between state $y^1$ and $y^2$, and equilibration rate $k^\mathrm{A}=2/(1/k_{12}^{A}+1/k_{21}^{A})$. Similarly, when the environment is in state $X=x^{B}$, the corresponding parameters are $\beta \Delta E^B_{12} = \ln \frac{k_{21}^{B}}{k_{12}^{B}}$ and $k^\mathrm{B}=2/(1/k_{12}^{B}+1/k_{21}^{B})$. Let $\Delta E^A=\lvert \Delta E^A_{12} \rvert$ and $\Delta E^B=\lvert \Delta E^B_{12} \rvert$ be the magnitudes of the energy gaps in environment states $x^A$ and $x^B$, respectively. 
The energy gaps $\Delta E^A$ and $\Delta E^B$ are free to be aligned ($\Delta E^A_{12} \Delta E^B_{12} > 0$) or anti-aligned ($\Delta E^A_{12} \Delta E^B_{12} < 0$).
The system's overall equilibration rate is thus
\begin{equation}
	\label{eq:ksys_unweighted}
	k_\mathrm{sys}=\frac{2}{\frac{1}{k^A}+\frac{1}{k^B}} \ . 
\end{equation}

\begin{figure}[H]
\centering
\includegraphics[width=\textwidth]{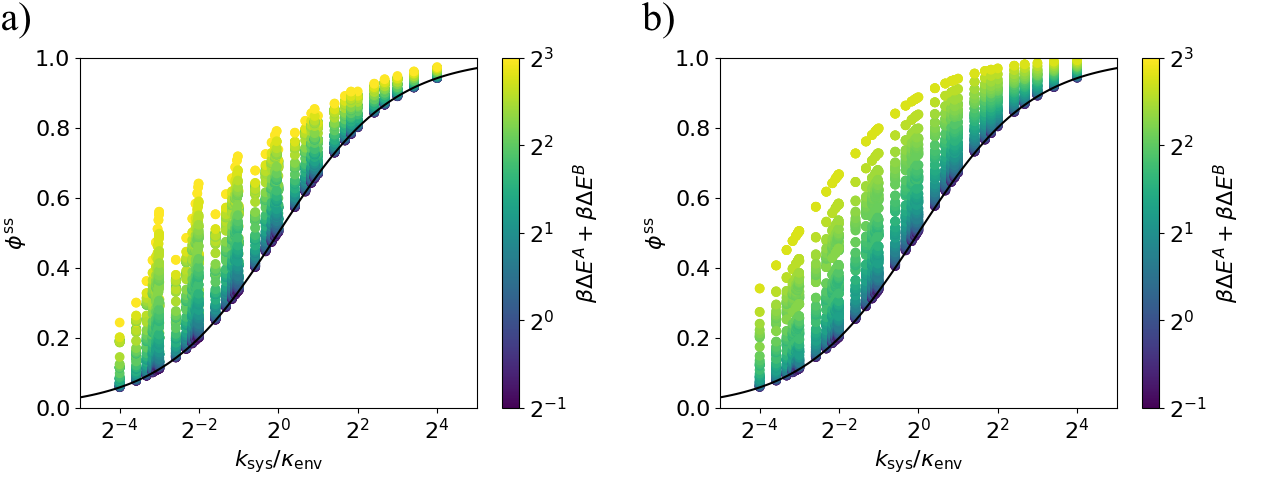}
\caption{{\bf Lower bound on dissipation ratio for fixed environment transition rate.}
The steady-state dissipation ratio $\phiSS$ is lower-bounded by the black curve~\eqref{eq:bound} for all values of the transition ratio $\kstar$. Each point corresponds to a particular set of parameters $k^A$, $k^B$, $\beta \Delta E^A$, and $\beta \Delta E^B$. (\textbf{a}) Models in which the energy gaps $\Delta E^A$ and $\Delta E^B$ are anti-aligned. (\textbf{b}) Models in which the energy gaps $\Delta E^A$ and $\Delta E^B$ are aligned. ($\kenv\Delta t = 10^{-12}$.)}
\label{fig:phi_bound}
\end{figure}

Equations~\eqref{eq:limit} and \eqref{eq:limit_smallprobs} also apply in this case of arbitrary system rates.
Figure~\ref{fig:phi_bound} shows that across the explored parameter space, the steady-state dissipation ratio $\phiSS$ lies above the bound~\eqref{eq:bound}, with $\phiSS$ approaching the bound in the weak-driving limit, $\beta (\Delta E^A+\Delta E^B) \ll 1$. 
We conclude that Eqs.~\eqref{eq:bound} and \eqref{eq:bound_smallprobs} apply for arbitrary system rates.

\subsection{Arbitrary Environment Rates}
Here we generalize our previous assumption of a fixed environmental transition rate $\kenv$, independent of the present environmental state.
We now allow for two different transition rates, $\kappa^{AB}$ and $\kappa^{BA}$, out of the two states $A$ and $B$ (Fig. \ref{fig:statemaps}c).

As above, we define the system equilibration rate $k^A$ and $k^B$ when the environment is in states $X=x^A$ and $X=x^B$, respectively. The overall system equilibration rate is the harmonic mean of the system transition rates for each environment state, weighted by the steady-state probabilities
\begin{equation}
	\label{eq:ksys}
    \ksys=\frac{1}{\frac{p^{\rm ss}(x^A)}{k^{A}}+\frac{p^{\rm ss}(x^B)}{k^B}} \ .
\end{equation}
For a uniform environmental transition rate (independent of environment state), this reduces to the previous un-weighted harmonic mean~\eqref{eq:ksys_unweighted}. 
Here we define $\kenvAv$ as the arithmetic mean of the transition rates between the environment states
\begin{equation}
	\label{eq:kenv}
\kenvAv=\frac{\kappa^{AB}+\kappa^{BA}}{2}. \
\end{equation}

\begin{figure}[H]
\centering
\includegraphics[width=\textwidth]{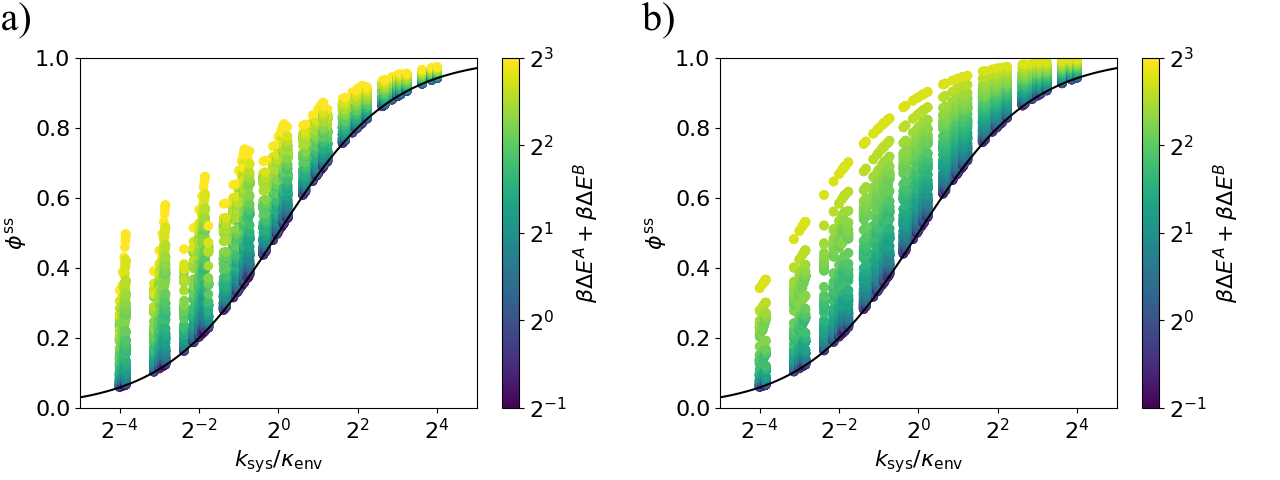}
\caption{{\bf Lower bound on dissipation ratio for varying environment transition rate.}
The steady-state dissipation ratio $\phiSS$ is lower-bounded by the black curve~\eqref{eq:bound} for all values of the transition ratio $\kstar$. Each point corresponds to a particular set of parameters $k^A$, $k^B$, $\beta \Delta E^A$, and $\beta \Delta E^B$. The environment transition rates are 
$\kappa^{AB}=1.8\kenvAv$ and $\kappa^{BA}=0.2\kenvAv$. (\textbf{a}) Models in which the energy gaps $\Delta E^A$ and $\Delta E^B$ are anti-aligned. (\textbf{b}) Models in which the energy gaps $\Delta E^A$ and $\Delta E^B$ are aligned. ($\kenvAv\Delta t = 10^{-12}$.)}
\label{fig:phi_bound_diffenv}
\end{figure}

With these definitions, Eqs.~\eqref{eq:limit} and \eqref{eq:limit_smallprobs} (replacing $\kenv$ with $\kenvAv$) apply to this case of arbitrary transition probabilities. Figure~\ref{fig:phi_bound_diffenv} shows that across a range of system and environment parameter values, bounds \eqref{eq:bound} and \eqref{eq:bound_smallprobs} hold.
The proposed bound depends on the system only through $\ksys$, and hence the environmental-state-dependent equilibration rates $k^A, k^B$.

\subsection{Beyond Two System and Environment States}
The expressions above suggest natural generalizations beyond the two environment and system states studied here. 
The rate at which the system responds to the environment is the overall system equilibration rate, which equals the 
harmonic mean 
of system transition rates given a particular environment state (because equilibration is a `series' process, with time scales [i.e., reciprocals of rates] adding),
weighted by the 
steady-state probability distribution over environment states. 
Therefore Eq.~\eqref{eq:ksys} can be written as 
\begin{equation}
	\label{eq:ksys_general}
	\ksys=\left[\sum_{i \in \mathcal{X}}\frac{p^\mathrm{ss}(x^i)}{k^{i}}\right]^{-1} \ .
\end{equation}
This definition extends to additional states simply by summing over the entire environment state space. 

The definition of $\kenvAv$ in Eq.~\eqref{eq:kenv} can also be generalized to additional states. 
The speed of environmental driving is determined by the overall rate of environmental change, which equals the 
arithmetic mean of transition rates out of environment states (because environmental change is a `parallel' process, with more transitions increasing the rate of change).
For total transition rate $\kappa_\mathrm{out}^{i}$ out of state $i \in \mathcal{X}$, in the simple two-state system $\kappa_\mathrm{out}^A=\kappa^{AB}$ and $\kappa_\mathrm{out}^B=\kappa^{BA}$. Thus, the rate in Eq.~\eqref{eq:kenv} can be written as 
\begin{equation}
	\label{eq:kenv_general}
\kenvAv=\frac{1}{N_\mathcal{X}}\sum_{i\in \mathcal{X}} \kappa_\mathrm{out}^{i} \ ,
\end{equation}
for $N_\mathcal{X}=\sum\limits_{i\in \mathcal{X}} 1$ distinct environment states. 
Since each of the quantities in the bound~\eqref{eq:bound_smallprobs} (with $\kenvAv$ replacing $\kenv$) are well-defined outside of the simple two-state system and environment studied here, it is 
intuitive
that this bound 
should
generalize as well.



\section{Discussion}

Ref.~\cite{Still2012} described a relationship between dissipation and nostalgia, a novel abstract information-theoretical concept quantifying the information the system stores about its environment that fails to be predictive of future environmental states. Energetically efficient performance requires avoiding this nostalgia. This framework suggests applications in biology, where living things are influenced by, and thus learn about, their environments. 
Recent explorations of the implications of this relationship have illuminated its behavior in model neurons~\cite{mcintosh2012information}, its relation to sensor performance~\cite{hartich2016sensory}, and the variation of it and related quantities across several biophysical model systems~\cite{Brittain2017}.

Here we focused on a
physical understanding of the relationships between the information-theoretic and thermodynamic quantities. We calculated the nostalgia in some model systems, alongside other thermodynamic and information-theoretic quantities of interest. 
Calculating these quantities over the parameter space of simple systems helps to establish an intuitive picture: 
when the system is quick to relax and strongly driven by the environment (energy gaps vary strongly with environment state), the nostalgia provides a tight lower bound on the steady-state dissipation~\eqref{eq:limit_smallprobs}; equivalently, the system learns more about the environment per unit heat dissipated. 

For fixed equilibration rates, we found the ratio of nostalgia to total dissipation is minimized in the weak-driving limit. Further, the ratio of nostalgia to total dissipation is bounded from below by this weak-driving limit~\eqref{eq:bound_smallprobs}, which depends on the system only through its overall equilibration rate. If the system is driven quasi-statically by its environment, this bound dictates that the predictive inefficiency (nostalgia) is responsible for all thermodynamic inefficiency (dissipation). 
Contexts further from the quasi-static limit can be further from saturating the bound in Eq.~\eqref{eq:ToPinequality}, and hence have a smaller relative contribution from model inefficiency. 

One could explore more complex models than the simple Markovian two-state systems and environments described here. 
One could expand the system to more states~\cite{Barato2014}, or expand the environmental behavior through additional states or non-Markovian dynamics, since this theoretical framework does not restrict the form of these transitions. 




\vspace{6pt} 

\acknowledgments{
The authors thank Volodymyr Polyakov (Physics and Technology, Igor Sikorsky Kyiv Polytechnic Institute) for insightful comments on the manuscript. 
This work is supported by a Natural Sciences and Engineering Research Council of Canada (NSERC) Discovery Grant (D.A.S.) and a Tier-II Canada Research Chair (D.A.S.).
}

\authorcontributions{
M.E.Q. and D.A.S. conceived and designed the study; M.E.Q. performed the analytic and numerical calculations; M.E.Q. analyzed the data; 
M.E.Q. and D.A.S. wrote the paper. 
}

\conflictofinterests{
The authors declare no conflict of interest. The funding sponsors had no role in the design of the study; in the collection, analyses, or interpretation of data; in the writing of the manuscript, and in the decision to publish the results.
} 




\bibliographystyle{mdpi}


\end{document}